 \documentclass[preprint,showpacs,preprintnumbers,amsmath,amssymb]{revtex4}

\usepackage{graphicx,color}
\usepackage{dcolumn}
\usepackage{bm}

\begin{document}

\preprint{APS/123-QED}

\title{Validity of the relativistic impulse approximation for elastic proton-nucleus
scattering at energies lower than 200~MeV}

\author{Z. P. Li$^{1}$}
\author{G.~C. Hillhouse$^{2,1}$}
\author{J. Meng$^{1,2,3,4}$}

\affiliation{$^{1}$School of Physics, Peking University, Beijing 100871}

\affiliation{$^{2}$Department of Physics, University of Stellenbosch, Stellenbosch, South Africa}

\affiliation{$^{3}$Institute of Theoretical Physics, Chinese Academy of Science, Beijing 100080}

\affiliation{$^{4}$Center of Theoretical Nuclear Physics, National
Laboratory of Heavy Ion Accelerator, Lanzhou 730000}

\date{\today}

\begin{abstract}
We present the first study to examine the validity of the relativistic impulse
approximation (RIA) for describing elastic proton-nucleus scattering at
incident laboratory kinetic energies lower than 200~MeV. For simplicity
we choose a $^{208}$Pb target, which is a spin-saturated spherical nucleus
for which reliable nuclear structure models exist. Microscopic scalar and
vector optical potentials are generated by folding invariant scalar and
vector scattering nucleon-nucleon (NN) amplitudes, based on our recently
developed relativistic meson-exchange model, with Lorentz scalar and vector
densities resulting from the accurately calibrated PK1 relativistic
mean field model of nuclear structure. It is seen that phenomenological Pauli
blocking (PB) effects and density-dependent corrections to $\sigma$N and $\omega$N
meson-nucleon coupling constants modify the RIA microscopic scalar and vector optical potentials
so as to provide a consistent and quantitative description of all elastic scattering
observables, namely total reaction cross sections, differential cross sections,
analyzing powers and spin rotation functions. In particular, the effect of PB
becomes more significant at energies lower than 200~MeV, whereas phenomenological
density-dependent corrections to the NN interaction {\it also} play an increasingly
important role at energies lower than 100~MeV.
\end{abstract}


\maketitle

\section{\label{sec:intro}Introduction}

The relativistic impulse approximation (RIA) provides an excellent
quantitative description of complete sets of elastic proton
scattering observables from various spin-saturated spherical nuclei
at incident energies ranging from 200 to 400~MeV \cite{Mu87}. The
latter represents the theoretical framework for generating complex
microscopic optical potentials for solving the elastic scattering
Dirac equation. The reliability of the RIA has also been
demonstrated by the fact that, for the above energies, predictions
of elastic proton scattering observables are very similar to the
corresponding results based on the highly successful global Dirac
phenomenological optical potentials, which have been calibrated to
provide excellent quantitative predictions of elastic proton
scattering observables from stable nuclei ranging from $^{12}$C to
$^{208}$Pb and for incident energies between 20 and 1040~MeV
\cite{Co93}.

The Melbourne group have already developed a highly predictive
microscopic Schr\"{o}dinger model for describing elastic scattering at
energies between 20 and 800~MeV \cite{De00a,De00b}. Here, however,
we focus on the RIA which is based on the relativistic Dirac equation,
and is more attractive in the sense that the microscopic scalar
and vector optical potentials -- and consequently the corresponding
Schr\"{o}dinger-equivalent central and spin-orbit optical potentials
\cite{Mu87}-- are directly related to the Lorentz properties of the
mesons mediating the strong nuclear force, a connection lacking
in nonrelativistic models.

Recently we developed an energy-dependent Lorentz covariant
parameterization of the on-shell NN scattering matrix at incident
laboratory kinetic energies ranging from 40 to 300~MeV \cite{Li08}.
In this paper, we employ the latter meson-exchange model to generate
relativistic microscopic scalar and vector folding optical
potentials in order to systematically study the predictive power of
the RIA for describing elastic proton-nucleus scattering observables
at energies lower than 200~MeV. At these low energies of interest,
multiple scattering effects \cite{Ch95}, medium modifications of the
NN interaction \cite{De00a}, and Pauli blocking \cite{Mu87,St05}
contributions become increasingly important, and hence the validity
of the RIA needs to be investigated at these low energies.

In Sec.~(\ref{sec:form}), we briefly describe the RIA as well as the
relevant input pertaining to our application thereof. Results are
presented in Sec.~(\ref{sec:results}), and we summarize and conclude
in Sec.~(\ref{sec:summary and conclusions}).

\section{\label{sec:form}Formalism}

\subsection{\label{subsec:ria} Relativistic Impulse Approximation
(RIA)}

A comprehensive description of the RIA, as well as the corresponding
computer codes and associated numerical details can be found in
Refs.~\cite{Mu87} and \cite{Ho91}. Here, we just give a brief
introduction of it. The Dirac optical potential is given by
\begin{equation}
  \label{eq:Uopt}
  U_{\rm opt}(q,E)=-\frac{4\pi i p}{M}\langle \Psi|\sum\limits^A_{n=1}
                e^{i{\bf q}\cdot{\bf r}(n)}\hat{\cal F}(q,E;n)|\Psi\rangle
\end{equation}
where $p$ is the magnitude of the three-momentum of the projectile in
the nucleon-nucleus center-of-mass frame, $|\Psi\rangle$ is the
$A$-particle ground state, and $\hat{\cal F}$ is the nucleon-nucleon
scattering operator. In the original RIA, $\hat{\cal F}$ is chosen
as:
\begin{equation}
  \label{eq:F}
  \hat{\cal F}(q,E)=\sum\limits_L F^L(q,E)\lambda^L_{(1)}\cdot\lambda^L_{(2)}
\end{equation}
where scattering amplitudes $F^L(q,E)$ are complex functions of the
momentum transfer $q$ and laboratory energy $E$, which can be
obtained by the relativistic Horowitz-Love-Franey model (see section
\ref{subsec:HLF}), and $\lambda^L_{(i)}$ stand for five Dirac
matrices for the projectile or target nucleon.

Assuming the ground state is a Hartree product of single-particle
4-component wave functions $U_\alpha({\bf r})$, the action of the
optical potential on the incident wave $|U_0\rangle$, projected onto
coordinate space, can be written as:
\begin{eqnarray}
  \label{eq:opt}
  \langle{\bf r}|U_{\rm opt}(q,E)|U_0({\bf r})\rangle &=& -\frac{4\pi i p}{M}
         \sum\limits_L\left[\int {\rm d}^3r^\prime \rho^L({\bf r^\prime})
         t^L_D(|{\bf r^\prime}-{\bf r}|;E)\right]\lambda^L U_0({\bf r})\nonumber\\
         & & -\frac{4\pi i p}{M}
         \sum\limits_L\left[\int {\rm d}^3r^\prime \rho^L({\bf r^\prime},{\bf r})
         t^L_X(|{\bf r^\prime}-{\bf r}|;E)\right]\lambda^L U_0({\bf r^\prime})
\end{eqnarray}
where
\begin{equation}
  t^L_{D}(|{\bf r}|;E)\equiv\int\frac{{\rm d}^3 q}{(2\pi)^3}t^L_{D}(q,E)e^{-i{\bf q}\cdot{\bf r}}
\end{equation}
with $t^L_D(q,E)\equiv(iM^2/2E_ck_c)F^L_D(q)$ and similarly for the
exchange pieces $t^L_X(Q,E)$. The nuclear densities are defined as:
\begin{equation}
  \rho^L({\bf r^\prime},{\bf r})\equiv\sum\limits^{occ^\prime}_i
  \bar U_i({\bf r^\prime})\lambda^L U_i({\bf r}),\ \ \ \ \ \
  \rho^L({\bf r})\equiv\rho^L({\bf r},{\bf r})
\end{equation}
Here the prime on the occupied states means that one sums over
target protons when the density is to be used with $pp$ amplitudes
and over target neutrons when it is to be used with $pn$ amplitudes.

The first term in Eq.~(\ref{eq:opt}) contains a multiplicative
factor that defines the direct optical potential:
\begin{equation}
  U^L_D(r,E)= -\frac{4\pi i p}{M}\int {\rm d}^3r^\prime \rho^L({\bf r^\prime})
                   t^L_D(|{\bf r^\prime}-{\bf r}|;E)
\end{equation}
Adopting a local-density approximation, the second term can be
localized to give:
\begin{equation}
  U^L_X(r,E)= -\frac{4\pi i p}{M}\int {\rm d}^3r^\prime \rho^L({\bf r^\prime},{\bf r})
                   t^L_X(|{\bf r^\prime}-{\bf r}|;E)j_0(p|{\bf r^\prime}-{\bf r}|)
\end{equation}
where $j_0$ is a spherical Bessel function. For spin-zero nucleus,
the non-zero densities are only baryon, scalar and a tensor term
associated with $\sigma^{0i}$. Hence, the RIA optical potential can
be written as
\begin{equation}
  \label{eq:UI=0}
  U_{\rm opt}(r;E)=U^S(r;E)+\gamma^0U^V(r;E)-2i{\mathbf{\alpha}}\cdot\hat{\bf r}U^T(r;E)
\end{equation}
where
\begin{equation}
  U^L(r;E)=U^L_D(r;E)+U^L_X(r;E)
\end{equation}
These optical potentials serve as input to solve the Dirac equation
for elastic proton scattering so as to generate the relevant
partial-wave scattering phase shifts for computing the scattering
observables, namely the total reaction cross section $\sigma_{R}$,
the differential cross section $d \sigma / d \Omega$, analyzing
power $A_{y}$ and spin rotation function $Q$. Similar to
Ref.~\cite{Mu87}, we resolve ambiguities in the form of the
relativistic NN scattering amplitudes by employing pseudovector
coupling for the $\pi$NN vertex. In Ref.~\cite{Mu87}, Horowitz and
Murdock have shown that the tensor potential has a negligible effect
on all the observables and nuclei of interest. In a similar vain and
for comparison to the latter, we also neglect the small tensor term.

\subsection{\label{subsec:HLF} Relativistic Horowitz-Love-Franey model (HLF)}

The original Love-Franey model and its relativistic version, i.e.,
Horowitz-Love-Franey model, have been described in detail in
Ref.~\cite{Lo81,Fr85,Ho85,Li08} and references therein. Here, we
briefly allude to list some important formula and the fitting
procedure at lower energy. Essentially HLF model parameterizes the
complex relativistic amplitudes $F^{L}(q,E)$ in terms of a set of $N
= 10$ meson exchanges in first-order Born approximation, such that
both direct and exchange NN (tree-level) diagrams are considered
separately, that is:
\begin{eqnarray}
F^{L}(q,E) = \frac{iM^2}{2 E_c k_c}[F^{L}_{D}(q) + F^{L}_{X}(Q)] \;,
\label{e-dplusex}
\end{eqnarray}
where
\begin{eqnarray}
 \label{e-fidirect}
 F^{L}_{D}(q) &=& \sum_{i=1}^{N}\delta_{L, L(i)}
  \langle \vec{\tau}_{1} \cdot \vec{\tau}_{2} \rangle^{T_i}
 f^{i}(q)\\
 \label{e-fierzexchange}
 F^{L}_{X}(Q) &=&
  (-1)^{T_{NN}}\sum_{i=1}^{N}C_{L(i), L} \langle \vec{\tau}_{1} \cdot
  \vec{\tau}_{2} \rangle^{T_i} f^{i}(Q)\;.
\end{eqnarray}
Here $T_i$ = (0,1) denotes the isospin of the $i^{\mbox{th}}$ meson,
$T_{NN}$ refers to the total isospin of the two-nucleon system,
$\langle \vec{\tau}_{1} \cdot \vec{\tau}_{2} \rangle^{T_i}$ is 1 or
-3 for different $T_{NN}$ and $T_i$, $C_{L(i), L}$ is the Fierz
matrix \cite{Fi37}, and
\begin{eqnarray}
\label{eq:fix}
f^{i}(x) = \frac{g_{i}^{2}}{{x}^{2} + m_{i}^{2}}(1 +
\frac{{x}^2}{\Lambda_i^2})^{-2} - i\frac{\bar{g}_{i}^{2}}{{x}^{2} +
\bar{m}_{i}^{2}}(1 + \frac{{x}^2}{\bar{\Lambda}_i^2})^{-2}
\end{eqnarray}
where $x$ represents the magnitude of either the direct
three-momentum transfer $q$ or the exchange-momentum transfer $Q$,
$(g_{i}^{2},\bar{g}_{i}^{2})$, $(m_{i},\bar{m}_{i})$, and
$(\Lambda_i,\bar{\Lambda}_i)$ are the real and imaginary parts of
the coupling constant, mass, and cutoff parameter for the
$i^{\mbox{th}}$ meson. These parameters are obtained by fitting the
theoretical amplitudes with the values extracted from the NN
scattering data.

Recently we have developed a set of parameter at incident laboratory
kinetic energies ranging from 40 to 300 MeV \cite{Li08}, in which
the coupling constant is set as energy dependent, namely
\begin{equation}
  g^{2}(E)\ =\ g^{2}_{0}[1\ +\ a_g(e^{a_{T} T_{\rm{rel}}}\ -\ 1)]
  \label{e-expfunc}
\end{equation}
where
\begin{equation}
  T_{\rm{rel}}\equiv \frac{T_0-T_{\rm{lab}}}{T_0}
\end{equation}
with $T_{0} = 200$~MeV, $T_{\rm{rel}}$ is positive in the 50 to
200~MeV energy range of interest, and $g^{2}_{0}$, $a_{g}$ and
$a_{T}$ are dimensionless parameters extracted by fitting to the
relevant data.

\subsection{\label{subsec:RMF} Relativistic Mean Field theory (RMF)}

For extracting the relevant scalar and vector proton and neutron
densities to be folded with the HLF NN scattering matric, we employ
the so-called PK1 Lagrangian density \cite{Lo04}, namely
\begin{eqnarray}
  \label{lagrangian}
   \displaystyle
   {\cal L}
     & = &
      \bar\psi\left[i\gamma^\mu\partial_\mu - M
            - g_\sigma\sigma
            - g_\omega\gamma^\mu\omega_\mu
            - g_\rho\gamma^\mu\vec\tau\cdot\vec\rho_\mu
            - e \gamma^\mu A_\mu\frac{1-\tau_3}{2}\right]\psi \nonumber\\
     &   & \mbox{}
       + \frac{1}{2} \partial_\mu \sigma \partial^\mu \sigma
       - \frac{1}{2} m_\sigma^2\sigma^2-\frac{1}{3}g_2\sigma^3-\frac{1}{4}g_3\sigma^4
    \nonumber \\
   &   & \mbox{}
    - \frac{1}{4} \Omega_{\mu\nu} \Omega^{\mu\nu}
    + \frac{1}{2} m_\omega^2 \omega_\mu \omega^\mu+\frac{1}{4}c_3(\omega_\mu \omega^\mu)^2
   \nonumber \\
   &   & \mbox{}
    - \frac{1}{4} \vec R_{\mu\nu} \vec R^{\mu\nu}
    + \frac{1}{2} m_{\rho}^{2} {\vec \rho}_\mu\cdot{\vec\rho}^\mu
   \nonumber \\
   &   & \mbox{}
    - \frac{1}{4} F_{\mu\nu} F^{\mu\nu},
\end{eqnarray}
where the parameters are listed in Table \ref{tab:PK1}, and the field
tensors for the vector mesons and the photon are respectively
defined as,
\begin{eqnarray}
 \displaystyle
 \left\{
  \begin{array}{rcl}
   \Omega_{\mu\nu}  & = & \partial_\mu\omega_\nu
                         -\partial_\nu\omega_\mu, \\
   \vec R_{\mu\nu} & = & \partial_\mu\vec \rho_\nu
                         -\partial_\nu\vec \rho_\mu, \\
   F_{\mu\nu}       & = & \partial_\mu {A}_\nu
                         - \partial_\nu {A}_\mu.
  \end{array}
  \right.
  \label{eq:tensors}
\end{eqnarray}

\begin{table}[htb]
\caption{\label{tab:PK1} The PK1 effective interaction. The masses
(in MeV), meson-nucleon couplings, and nonlinear coefficients are
listed.}
\begin{tabular}{ccccccccccc}
\hline\hline
$M_n$&$M_p$&$m_\sigma$&$m_\omega$&$m_\rho$&$g_\sigma$&$g_\omega$&$g_\rho$&$g_2$&$g_3$&$c_3$ \\
\hline
\ 939.5731\ &\ 938.2796\ &\ 514.0891\ &\ 784.254\ &\ 763\ &\ 10.3222\ &\ 13.0131\ &\ 4.5297\ &\ -8.1688\ &\ -9.9976\ &\ 55.636\\
\hline\hline
\end{tabular}
\end{table}
Under the variation of the Langrangian density with respect to the
different fields, one obtains the equations of motion for the
nucleon and different mesons via a self-consistent procedure. To
obtain the parameters, the masses of spherical nuclei $^{16}$O,
$^{40}$Ca, $^{48}$Ca, $^{56}$Ni, $^{68}$Ni, $^{90}$Zr, $^{116}$Sn,
$^{132}$Sn, $^{194}$Pb, and $^{208}$Pb, the compression modulus $K$,
the baryonic density at saturation $\rho_{\rm sat}$, and the
asymmetry energy $J$ of nuclear matter are fitted to give the
minimum by the Levenberg-Marquardt method. The relative deviations
are 0.0102 and 0.0178 for the masses of 19 spherical nuclei and
charge radii of 13 spherical nuclei, respectively, which are similar
or better than the other commonly used RMF effective interactions,
such as the NL1 \cite{Re86}, PL-40 \cite{Re88}, NL-SH \cite{Sh93},
TM1 \cite{Su94}, NL3 \cite{La97}, TW99 \cite{Ty99}, and DD-ME1
\cite{Ni02}. Compared to these effective interactions, the PK1
parametrization represents an improvement in the sense that
center-of-mass corrections are included microscopically and it also
provides a unified description of the properties of stable and
unstable nuclei over a broad mass ($16 \le A \le 214$) range.

\begin{figure}[htb]
\includegraphics[scale=1]{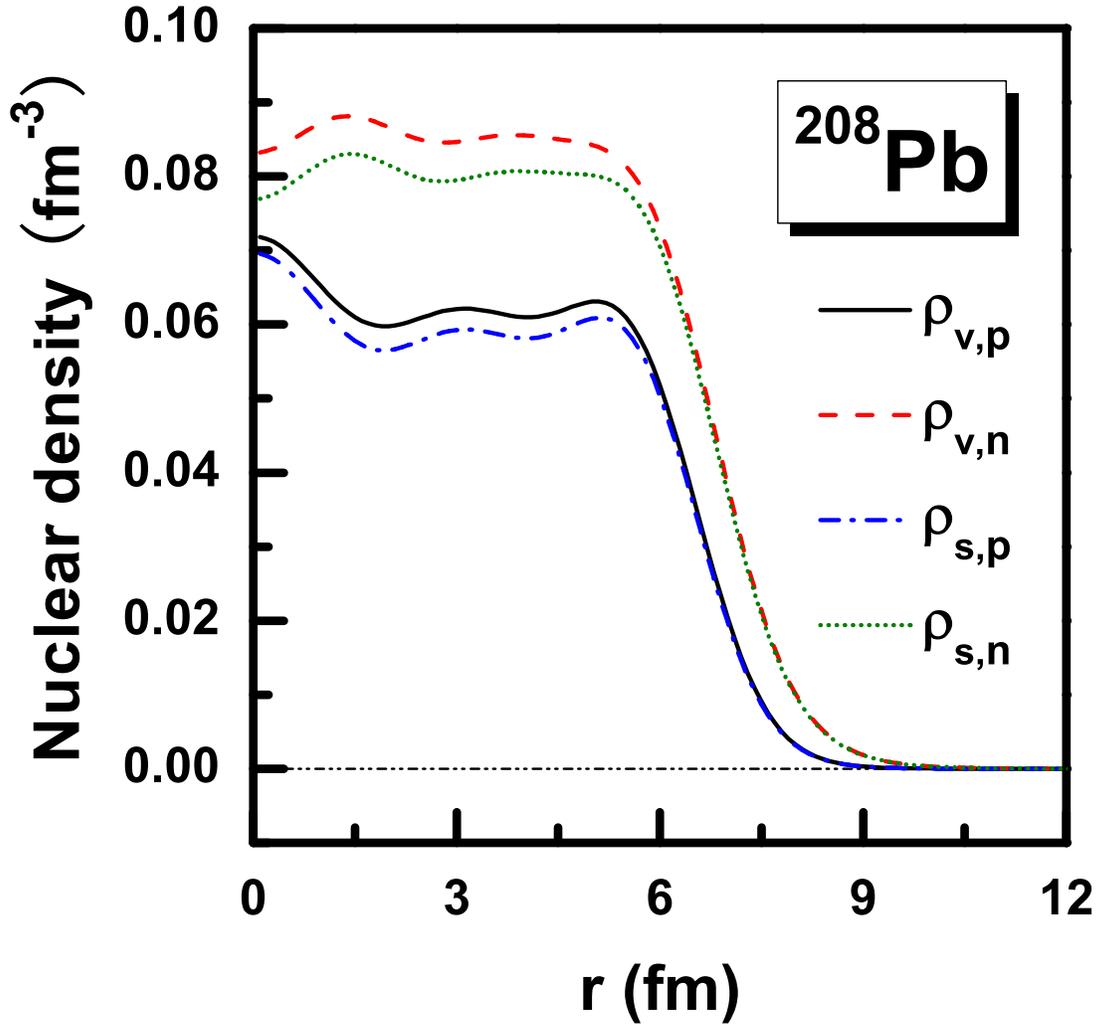}
\caption{\label{fig:rho} (Color online) Scalar, and vector (baryon)
density (in units of fm$^{-3}$) distribution in radial direction $r$
(in fm) for proton and neutron of $^{208}$Pb, respectively. The
proton and neutron densities are denoted by solid and dashed curves,
respectively. The dash dotted curve corresponds to the proton scalar
density, and dotted curve indicates the neutron scalar density. The
PK1 effective interaction is used.}
\end{figure}

At last, to minimize uncertainties associated with nuclear structure
input, we focus on proton scattering from $^{208}$Pb which is a
spin-zero spherical nucleus for which the PK1 effective interaction
can give excellent description. In Fig.~\ref{fig:rho}, the scalar
and vector density distributions for proton and neutron of
$^{208}$Pb are shown. Another advantage for considering such a heavy
nucleus as $^{208}$Pb, is that at the low energies of interest,
recoil corrections to the Dirac scattering equation are expected to
be small and hence can be neglected.

\section{\label{sec:results}Results}

Guided by the availability of experimental data at the lower, middle
and higher-energy regions of the 50 -- 200~MeV range, we focus
on incident energies of 65 \cite{Sa81}, 98 \cite{Na81}, 121 \cite{Sc82}
and 200~MeV \cite{Hu88,Ot88}, and examine the validity of our
RIA predictions for describing $d \sigma / d \Omega$, $A_{y}$ and $Q$
from a $^{208}$Pb target nucleus as the energy is systematically lowered.
Note that at approximately 100~MeV, there exists no complete set of
experimental data for one specific energy, and hence we compare our
RIA calculations (in the mid-energy range) to data for $d \sigma / d \Omega$
and spin observables, $A_{y}$ and $Q$, at 121~MeV and 98~MeV, respectively.

\begin{figure}[htb]
\includegraphics[scale=0.8]{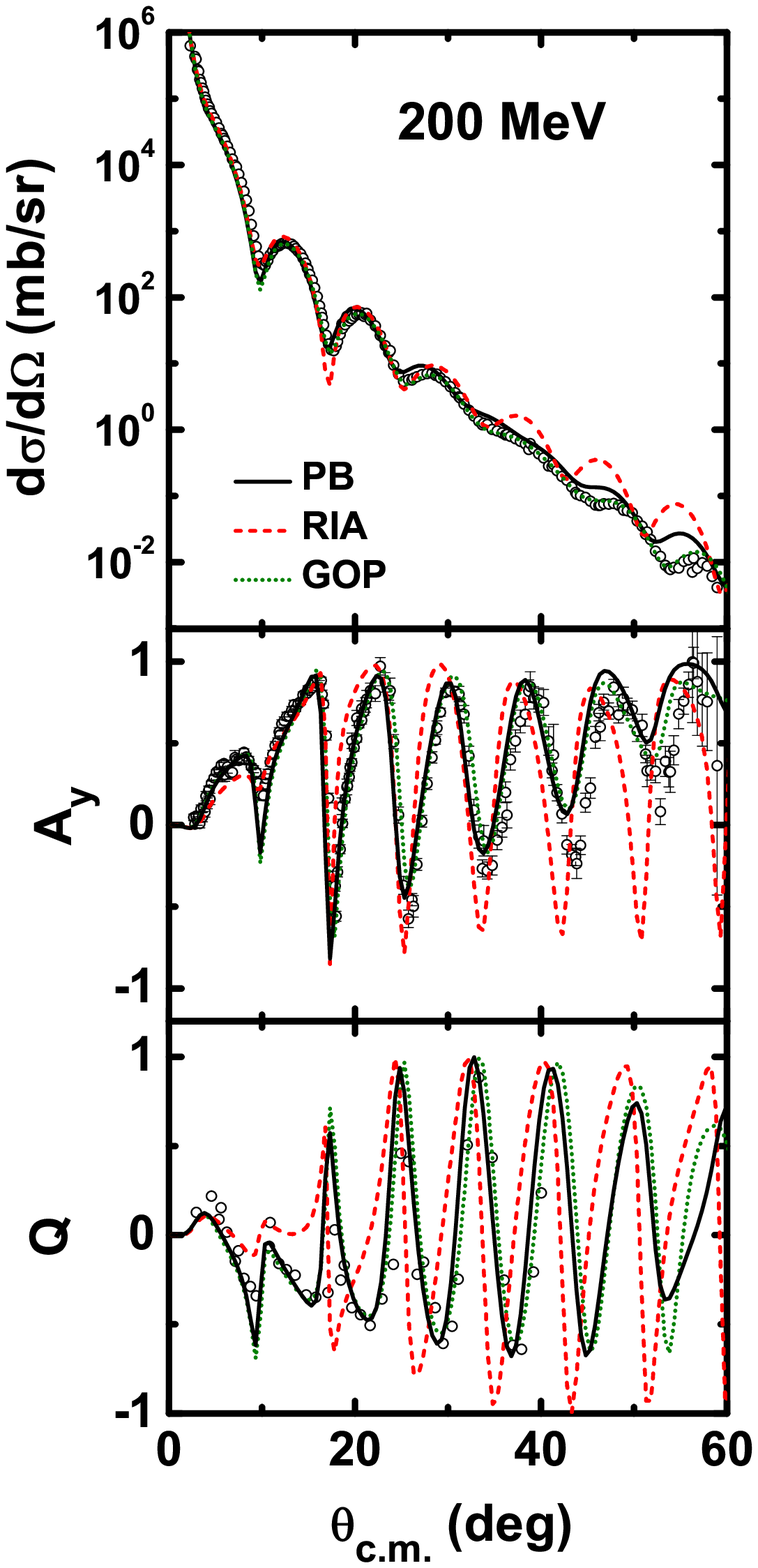}
\caption{\label{fig:obs200-Pb208} (Color online) Differential cross
sections $d\sigma/d\Omega$ (in units of $mb/sr$), analyzing powers
$A_y$ and spin rotation functions $Q$, plotted as a function of
center-of-mass scattering angles $\theta_{\rm{c.m.}}$ (in degrees),
for elastic proton scattering from $^{208}$Pb at incident laboratory
kinetic energy of 200~MeV. The experimental data \cite{Hu88,Ot88}
are denoted by open circles. Uncorrected RIA predictions are
represented by black dashed curves (legend: RIA; color red online),
the black dotted curves correspond to predictions based on the EDAI
Dirac global optical potentials (GOP) \cite{Co93} (legend: GOP;
color olive online), and RIA calculations including phenomenological
Pauli blocking (PB) corrections are indicated by the black solid
curves (legend: PB).}
\end{figure}

\begin{figure}[htb]
\includegraphics[scale=0.8]{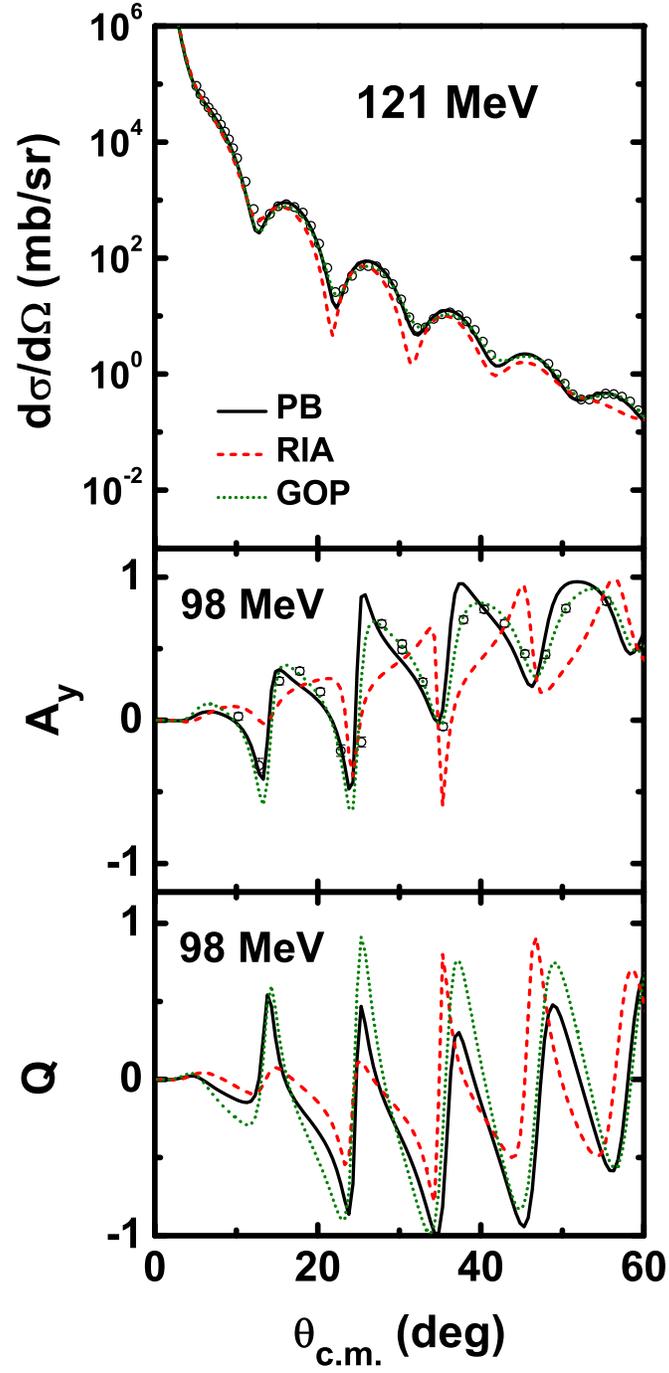}
\caption{\label{fig:obs121-Pb208} (Color online) Same as
Fig.~\ref{fig:obs200-Pb208}, except for incident laboratory
kinetic energies of 121 or 98~MeV. The experimental data at 121
\cite{Sc82} and 98 \cite{Na81} are denoted by open circles.
}
\end{figure}

\begin{figure}[htb]
\includegraphics[scale=0.8]{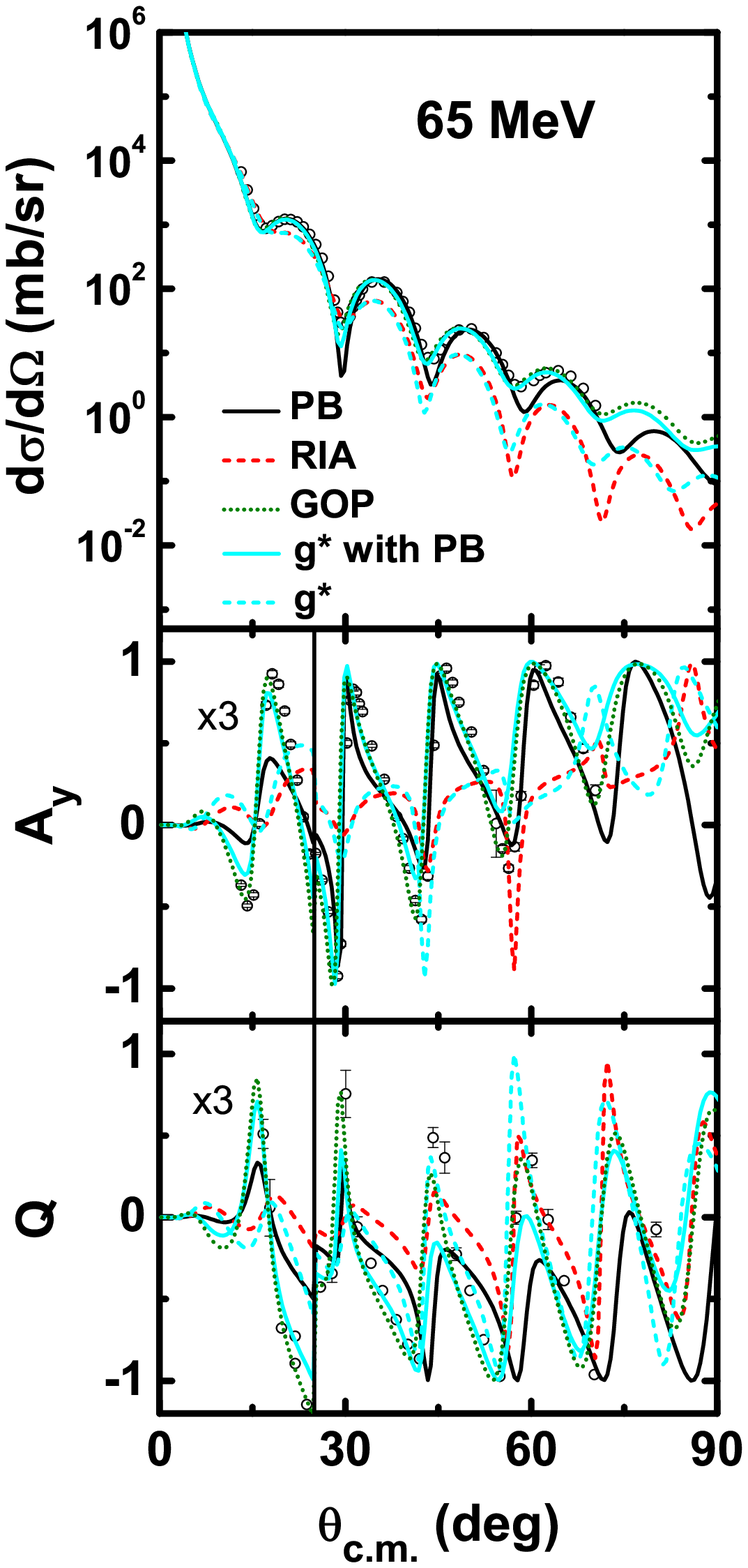}
\caption{\label{fig:obs65-Pb208} (Color online) Same as
Fig.~\ref{fig:obs200-Pb208}, except for an incident laboratory
kinetic energy of 65~MeV. The gray dashed curve (color cyan online)
corresponds to RIA predictions based on the renormalized
medium-modified $\sigma$N and $\omega$N coupling constants (denoted
by the legend: $g^{*}$), and the gray solid lines (color cyan
online) denote the combined effect of medium modified coupling
constants together with PB corrections (legend: $g^{*}$ with PB).
The experimental data \cite{Sa81} are denoted by open circles. The
analyzing powers $A_y$ and spin rotation functions $Q$ are
multiplied by 3 at the $\theta_{\rm{c.m.}}$ angles from 0 to 25
degrees.}
\end{figure}

In Figs.~\ref{fig:obs200-Pb208}, \ref{fig:obs121-Pb208}, and
\ref{fig:obs65-Pb208}, the RIA calculations are denoted by black
dashed curves (color red online) and for reference we also display
results based on the energy-dependent mass-independent (EDAI) global
Dirac optical potentials (GOP) \cite{Co93} indicated by the black
dotted curves (color olive online). We consider center-of-mass
scattering angles ranging up to a maximum value corresponding to a
three-momentum transfer of about $q \approx 2.5\ \rm{fm^{-1}}$ for
which first-order microscopic nonrelativistic models and
relativistic global optical potentials have been shown to be valid
\cite{De05}. As expected, the global Dirac optical potentials
provide an excellent description of all the observables for the
entire energy range of interest. At 200~MeV, the RIA provides a
satisfactory description of $d \sigma / d \Omega$ and $A_{y}$ for
angles ranging up to a value corresponding to the third minimum in
$d \sigma / d \Omega$. Although the RIA provides a good description
of $Q$ from 20 to 40 degrees, it fails to describe the small angle
behavior. At lower energies, the RIA provides a satisfactory
description of $d \sigma / d \Omega$ up to approximately the second
minimum, but fails to describe $A_{y}$ and $Q$.

\begin{figure}[htb]
\includegraphics[scale=0.8]{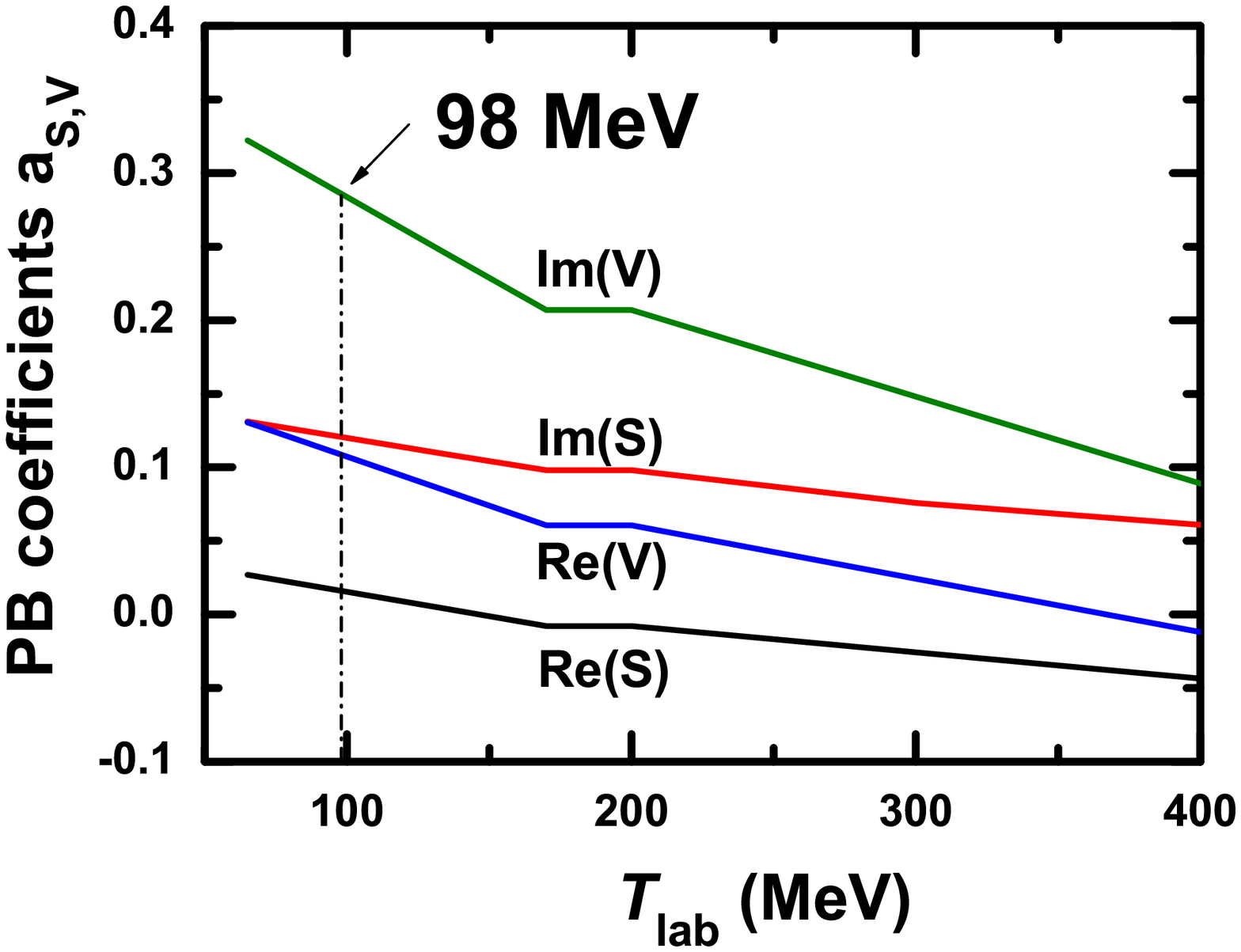}
\caption{\label{fig:aSV-T}(Color online) Linear interpolation and
extrapolation of PB correction factors calculated from a
relativistic Dirac-Brueckner approach. The four curves from top to
bottom indicate the PB correction factors corresponding to the
imaginary (Im) and real (Re) parts of the vector $V$ and scalar $S$
optical potentials. In the code corresponding to Ref. \cite{Ho91},
the values at $T_{\rm lab} = 170$~MeV are chosen same as $T_{\rm
lab} = 200$~MeV, which causes the flat section in the curves. The
linear extrapolation at $T_{\rm lab} = 98$~MeV is emphasized.}
\end{figure}

\begin{figure}[htb]
\includegraphics[scale=1.1]{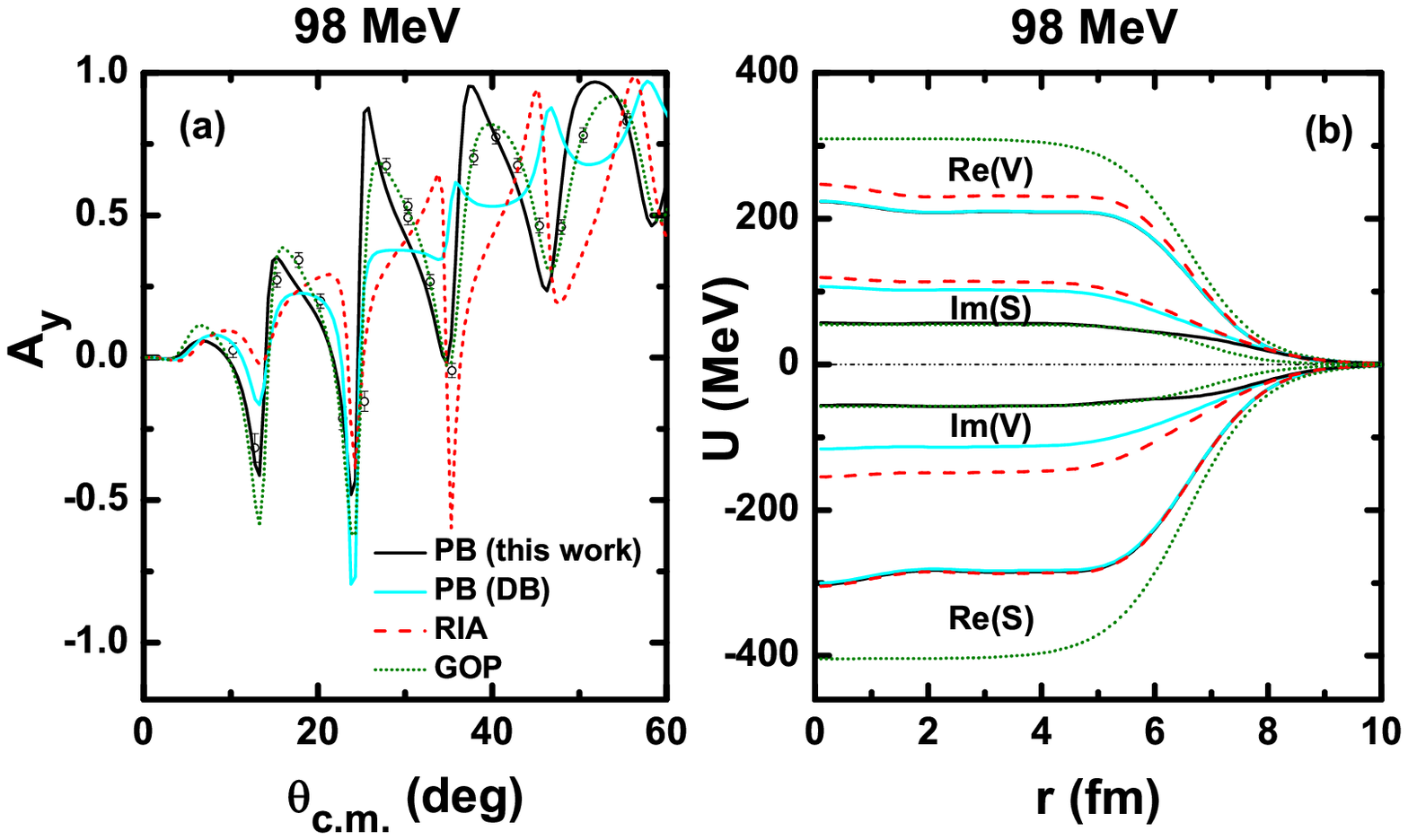}
\caption{\label{fig:Ay-pot-98}(Color online) (a) Analyzing power
$A_{y}$ for elastic proton scattering from $^{208}$Pb at 98~MeV,
plotted as a function of center-of-mass scattering angles
$\theta_{\rm{c.m.}}$ (in degrees). The experimental data are taken
from Ref.~\cite{Na81}. Uncorrected RIA predictions are represented
by black dashed curves (legend: RIA; color red online), the black
dotted curves (color olive online) correspond to predictions based
on the EDAI Dirac global optical potentials (GOP) \cite{Co93}
(legend: GOP), and RIA calculations including extrapolated
Dirac-Brueckner and phenomenological PB corrections are denoted by
gray [legend: PB (DB); color cyan online] and black solid curves
[legend: PB (this work)], respectively. (b) The corresponding real
(Re) and imaginary (Im) scalar $S$ and vector $V$ optical potentials
(in units of MeV) for elastic proton scattering from $^{208}$Pb at
98~MeV, plotted as a function of the nuclear radius $r$ (in units of
fm). In particular, the two black solid curves corresponding
to the real parts of the scalar and vector optical potentials are
covered by the gray curves.}
\end{figure}

For elastic proton scattering from a heavy nucleus such as $^{208}$Pb,
Murdock and Horowitz \cite{Mu87} have demonstrated that Pauli Blocking
(PB) corrections to the RIA optical potentials play an increasingly
significant role at lower energies. Essentially PB
represents the mechanism that prevents nucleons in the nuclear medium
from scattering to occupied intermediate states \cite{Ha70}. We now
proceed to study the effect of PB corrections for our low energy range
of interest. Following the procedure outlined in Ref.~\cite{Mu87},
we include PB corrections to the RIA microscopic optical potentials
via the so-called energy-dependent PB coefficients $a_{S,V}(T_{\rm{lab}})$,
that is,
\begin{eqnarray}
 U^{S,V}_{\rm{PB}}(r,T_{\rm{lab}}) & = & \left[1 - a_{S,V}(T_{\rm{lab}})
 \left(\frac{\rho_B(r)}{\rho_0}\right)^{2/3}\right]U^{S,V}(r,T_{\rm{lab}})
\end{eqnarray}
where $T_{\rm{lab}}$ denotes the incident laboratory kinetic energy,
$\rho_B(r)$ is the local RMF baryon density of the target nucleus,
$\rho_0=0.1934\ {\rm{fm}}^{-3}$, and $U^{S,V}(r,T_{\rm{lab}})$
represents the uncorrected RIA scalar $S$ and vector $V$ optical
potentials. Within the context of a relativistic Dirac-Brueckner
approach, values for the PB correction factors
$a_{S,V}(T_{\rm{lab}})$ have been extracted at discrete energies of
135, 200, 300 and 400~MeV \cite{Mu87} and also at discrete momenta
ranging from about 0.46 to 0.75~GeV/c \cite{Ha86}. Consistent with
the results reported in Ref.~\cite{Mu87}, in
Fig.~\ref{fig:obs200-Pb208} we confirm that PB corrections (black
solid curve) provide an improved description of the 200~MeV data,
the effect been most pronounced for $A_{y}$ and $Q$ at small angles.
Due to the enhanced sensitivity exhibited by spin observables, we
now focus on the influence of PB on $A_{y}$ at a lower energy of
98~MeV, where PB corrections are expected to be more significant:
note that there are no experimental data for $Q$ at 98~MeV. The lack
of published PB coefficients at this energy necessitates linear
extrapolation of the published Dirac-Brueckner values in
Ref.~\cite{Mu87}, which is shown in Fig.~\ref{fig:aSV-T}. RIA
predictions with and without PB corrections are denoted by the gray
solid (color cyan online) and black dashed (color red online) curves
in Fig.~\ref{fig:Ay-pot-98} (a), respectively. One clearly observes
that the extrapolated Dirac-Brueckner PB corrections cannot account
for the $A_{y}$ data.

Due to the latter failure, we now proceed to study PB via a
phenomenological approach. In particular, we study to which extent
phenomenologically extracted values of $a_{S,V}(T_{\rm{lab}})$ can
provide a consistent description of complete sets of scattering
observables at the specific energies of interest. Starting with the
lower momentum values of $a_{S,V}(T_{\rm{lab}})$ published in table
2 [$\rho_{0}$, $p$ = 0.46~GeV/c] of Ref.~\cite{Ha86}, we
systematically vary the real and imaginary parts to give one GOP
that provides the best fit-by-eye of $d \sigma / d \Omega$, $A_{y}$
and $Q$ for elastic proton scattering from $^{208}$Pb at five
discrete energies of 65, 100, 135, 170 and 200~MeV, spanning the
range of interest. In fact, this is enough for the purpose of this
paper: to check the validity of the RIA at lower energies. Our
phenomenological PB factors are listed in Table \ref{tab:PBa}
together with the corresponding published Dirac-Brueckner values
from Ref.~\cite{Mu87}.

\begin{table}[htb]
\caption{\label{tab:PBa} Phenomenological Pauli blocking correction
factors $a(T_{\rm{lab}})$ for real and imaginary RIA microscopic
scalar $S$ and vector $V$ optical potentials as various energies
$T_{\rm{lab}}$. Where available, the corresponding published
Dirac-Brueckner values \cite{Mu87} are indicated in square
brackets.}
\begin{tabular}{ccccc}
\hline\hline
 &\multicolumn{2}{c}{Scalar} & \multicolumn{2}{c}{Vector}\\
 Energy [MeV]  &  Real  &  Imaginary &  Real  & Imaginary  \\
\hline
65  & 0.020 &      0.68  &    0.14  & 0.85\\
100 & 0.007 &      0.60  &    0.11  & 0.73\\
135 & 0.005 [0.00377] &      0.53 [0.108825] &    0.09 [0.08403] & 0.60 [0.24535]\\
170 & 0.008 &      0.42  &    0.07  & 0.50\\
200 & 0.010 [-0.0078] &      0.35 [0.098] &  0.0605 [0.0605]  & 0.42 [0.207]\\
\hline\hline
\end{tabular}
\end{table}

We now consider the effect of these phenomenological corrections on
$A_{y}$ at 98~MeV, where the extrapolated Dirac-Brueckner
corrections were shown to fail. In Fig.~\ref{fig:Ay-pot-98} (a) one
clearly sees that our phenomenological PB corrections (black solid
curve) provide a more satisfactory description of $A_{y}$ compared
to the Dirac-Brueckner PB corrections (gray solid curve). Focusing
on the optical potentials, in Fig.~\ref{fig:Ay-pot-98} (b) we
observe that the main effect of both Dirac-Brueckner (grey solid
curves) and phenomenological (black solid curves) PB is to decrease
the strengths of the real and imaginary parts of the scalar and
vector optical potentials, the effect being most pronounced for the
imaginary parts. Furthermore, our phenomenological corrections have
a larger effect than the corresponding Dirac-Brueckner corrections.
Also note that the PB-corrected optical potentials differ in
strength from the corresponding global optical potentials (black
dotted curve), which suggests that the delicate interplay between
the relative strengths and signs of the scalar and vector optical
potentials, rather than absolute magnitudes, is responsible for
providing an accurate description of the diffractive behavior of all
scattering observables.

In Figs.~\ref{fig:obs200-Pb208}, \ref{fig:obs121-Pb208}, and
\ref{fig:obs65-Pb208}, we now display our theoretical predictions
including the above-mentioned phenomenological PB corrections (black
solid curves) over the entire range of interest. Although not shown,
both Dirac-Brueckner and phenomenological PB corrections provide the
same quantitative improvement for all observables at 200~MeV. At
121~MeV, PB corrections provide an excellent description of the
first few maxima and minima for $d \sigma /d \Omega$. There are no
data for $Q$ at 98~MeV, but we note that PB corrections give results
very similar to the corresponding global Dirac optical potentials.
At the lower energy of 65~MeV, PB corrections invoke a larger
effect and provide a satisfactory description of $d \sigma / d
\Omega$ and $A_{y}$ up to about 50 degrees, but predictions
deteriorate at larger angles. Note, however, that at 65~MeV, PB
corrections fail to quantitatively describe the first two extrema
for $A_{y}$, as well as fail to reproduce $Q$ over the entire
angular range. The 65~MeV results emphasize the important fact that
optical potentials can be deemed reliable only if complete sets of
elastic scattering observables, as opposed to only one or two
observables, are accurately described.

\begin{figure}[htb]
\includegraphics[scale=0.8]{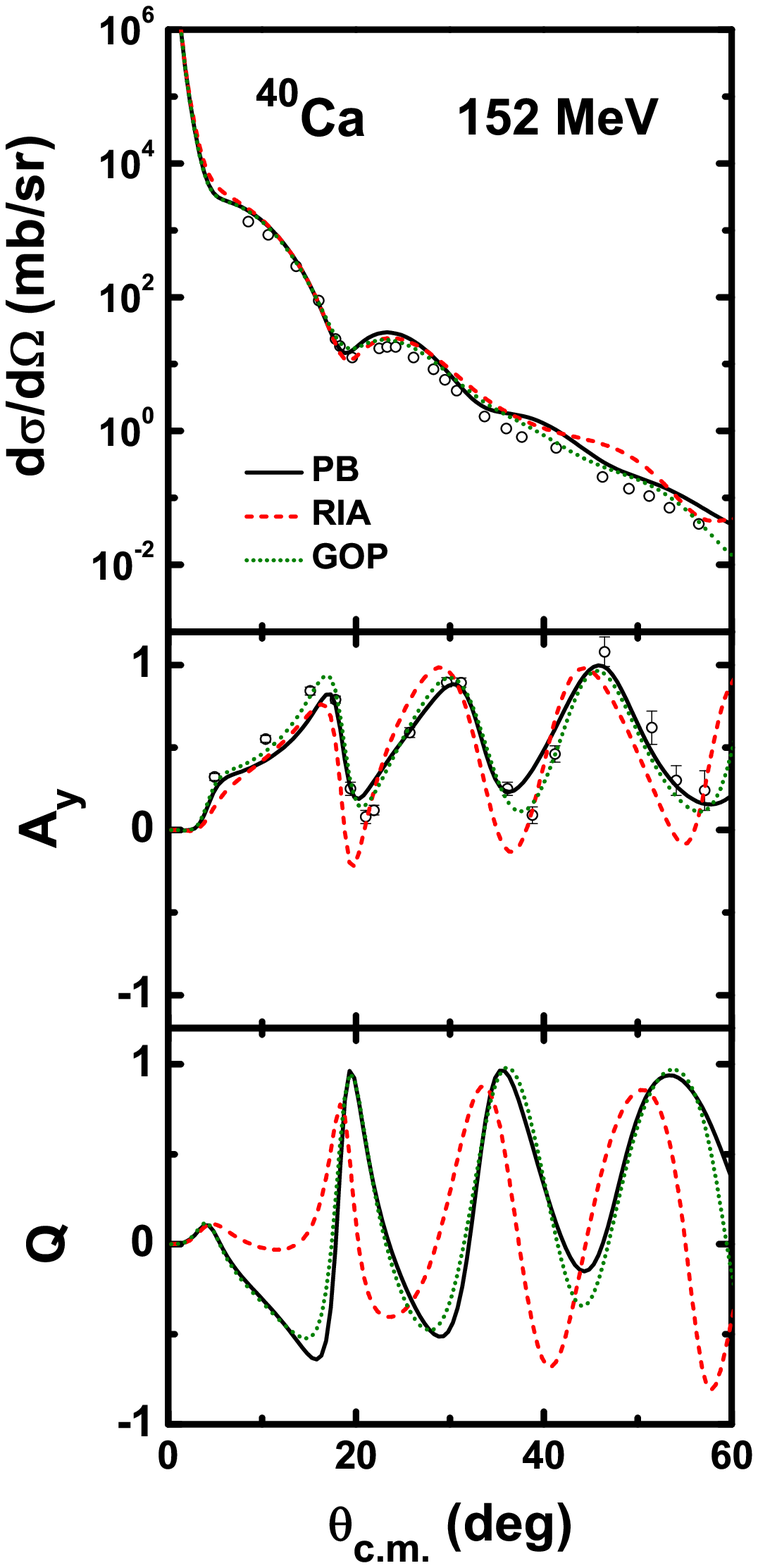}
\caption{\label{fig:obs2-Ca40} (Color online) Same as
Fig.~\ref{fig:obs200-Pb208}, except for elastic proton scattering
from $^{40}$Ca at incident laboratory kinetic energy of 152~MeV. The
experimental data at 152~MeV are taken from Ref.~\cite{Ro66}.}
\end{figure}

To further validate our phenomenological procedure for including PB
corrections, in Fig.~\ref{fig:obs2-Ca40} we now consider a different
spin-zero spherical target nucleus, namely $^{40}$Ca, at an incident
proton energy of 152~MeV: the meaning of the various curves is the
same as those in Fig.~\ref{fig:obs200-Pb208}. Compared to the
observables without PB (black dashed curves), and consistent with
our conclusions for $^{208}$Pb, the corrected values (black solid
curves) provide an improved description of all observables and also
agree with the GOP results.

\begin{figure}[htb]
\includegraphics[scale=1.4]{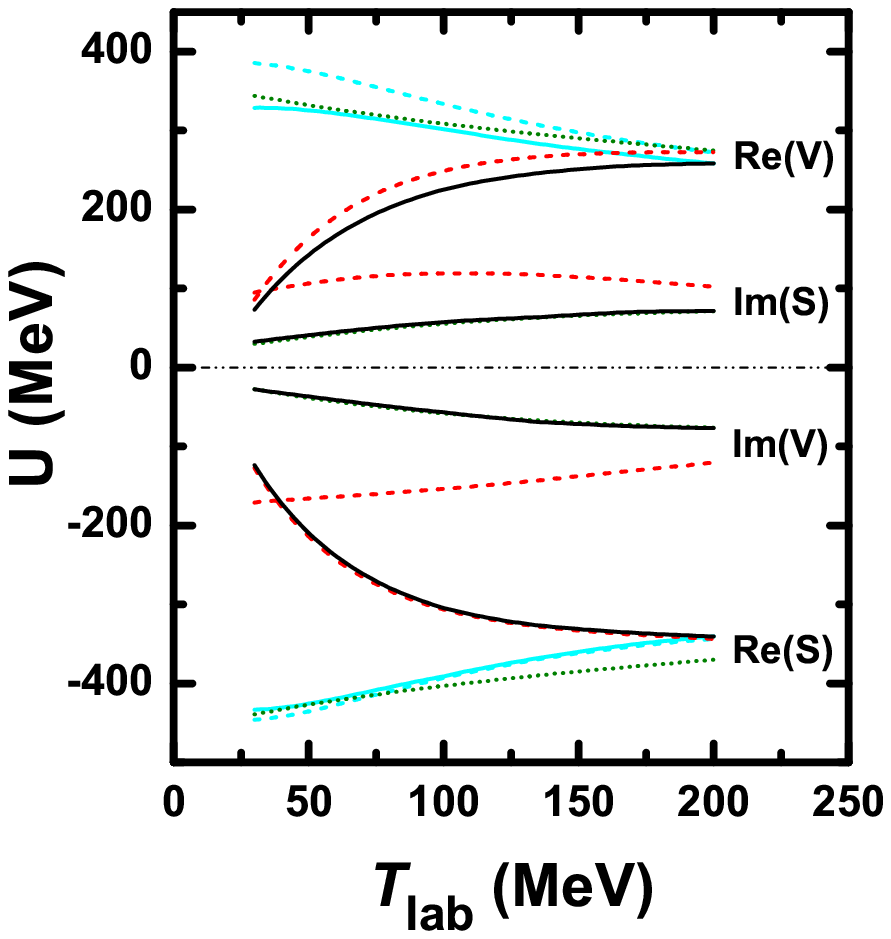}
\caption{\label{fig:pot-Tlab} (Color online) Strengths of the real
(Re) and imaginary (Im) scalar $S$ and vector $V$ optical potentials
(in units of MeV) at the center (radius $r = 0$) of a $^{208}$Pb
nucleus as a function of incident laboratory kinetic energies
$T_{\rm{lab}}$ (in units of MeV). The identification of the curves
is the same as in Fig.~\ref{fig:obs65-Pb208}.}
\end{figure}

Although PB systematically improves the RIA predictions at lower
energies, these corrections do not provide a quantitative
description of complete sets of observables, and hence it is clear
that other effects also begin to play an increasingly significant
role as the energy is lowered. The question now arises as to whether
one can incorporate additional corrections so as to consistently
improve these low energy results within the framework of the RIA.
Various authors \cite{Sa98,Hi03,De00b} have stressed that importance
of nuclear medium modifications to the NN interaction at lower
energies. In particular, a number of theoretical models
\cite{Se86,Br91,Fu92,Lo04} predict density-dependent corrections to
meson-nucleon coupling constants as well as nucleon- and
meson-masses in normal nuclear matter. We now adopt a
phenomenological approach to investigate to what extent the
renormalization of certain meson-nucleon coupling constants can
provide a systematic improvement of the data at 65~MeV. Since scalar
and vector RIA optical potentials are dominated by contributions
from $\sigma$- and $\omega$-meson exchange to the NN scattering
amplitudes, we vary the values of the real $\sigma$N and $\omega$N
coupling constants, $g^2_{\sigma}$ and $g^2_{\omega}$, so as to
provide the best consistent description of elastic proton-nucleus
scattering. The result of this procedure is that the value of
$g^2_{\sigma}$ changes from -7.5701 to -11.1734 and $g^2_{\omega}$
changes from 7.4511 to 12.9704: these new values of the couplings
correspond to $a_{g} = 0.0271$ and $a_T = 3.8364$ for the $\sigma$
meson and to $a_{g} = 0.1022$ and $a_T = 1.9527$ for the $\omega$
meson [see Eq.~(\ref{e-expfunc}) in this paper, also Eq.~(21) and
Table II in Ref.~\cite{Li08}]. The effect of renormalizing these
couplings is illustrated by comparing the original results (black
solid curves) to the renormalized values (gray solid curve) in
Fig.~\ref{fig:obs65-Pb208}. The most pronounced improvement is
observed for the minima of $d \sigma /d \Omega$ at large scattering
angles and spin observables at small scattering angles. At the level
of the optical potentials, the main effect of the modified coupling
constants is to increase the strengths of the real parts of the
scalar and vector potentials at lower energies such that the
energy-dependent trend these strengths is qualitatively similar to
the corresponding GOP results, as illustrated in
Fig.~\ref{fig:pot-Tlab}. Furthermore, these improved predictions
arising from renormalized coupling constants are consistent with
other nuclear reaction- and structure-studies focusing on
density-dependent corrections to the NN interaction
\cite{Sa98,Hi03,Lo04,Hi06}. Note, however, that the best description
of $A_{y}$ and $Q$ up to angles of about 40 degrees, is obtained by
including {\it both} PB and density-dependent corrections (gray
solid curve).

\begin{figure}[htb]
\includegraphics[scale=0.8]{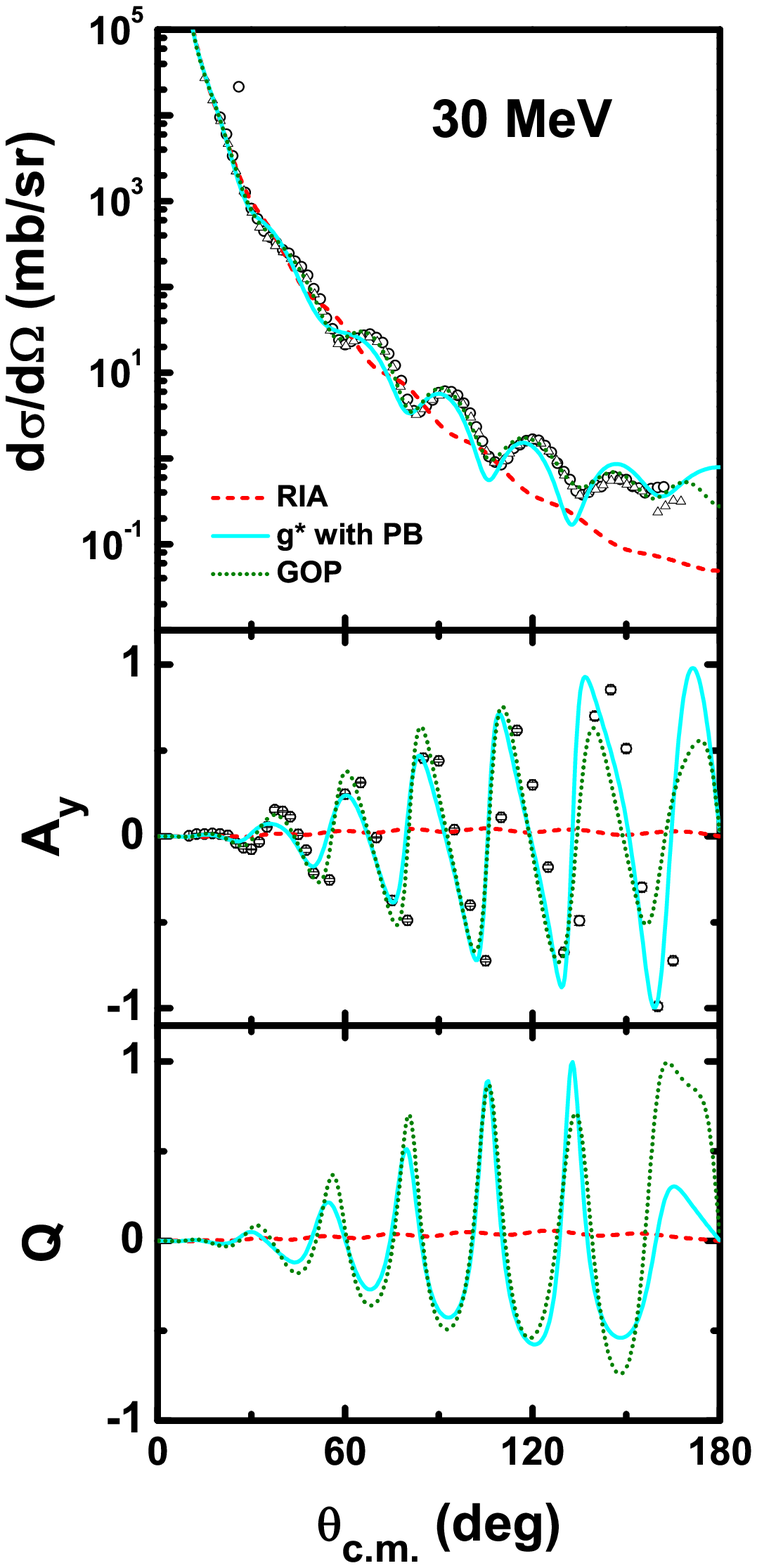}
\caption{\label{fig:obs30-Pb208}(Color online) Same as
Fig.~\ref{fig:obs65-Pb208}, except for an incident laboratory
kinetic energy of 30~MeV. The experimental data are taken from
Ref.~\cite{Oe74,Ri64}.}
\end{figure}

To further test the consistency of our scheme for including {\it both} PB and
density-dependent corrections, we now focus on elastic proton scattering
at an energy as low as 30~MeV where these effects are expected to be important.
In Fig.~\ref{fig:obs30-Pb208}, it is gratifying to observe that both
phenomenological corrections (gray solid curves) contribute significantly
toward providing an excellent description of the $d \sigma / d \Omega$ and $A_{y}$
data \cite{Oe74,Ri64} over the entire angular region. Although there are
no data for $Q$ at this energy, we note that our predictions are similar
to the GOP results which have constrained to fit elastic scattering observables
at energies as low as 20~MeV \cite{Co93}. The above mentioned results inspire
confidence in the predictive power of our phenomenological corrections to the RIA.

\begin{figure}[htb]
\includegraphics[scale=1.00]{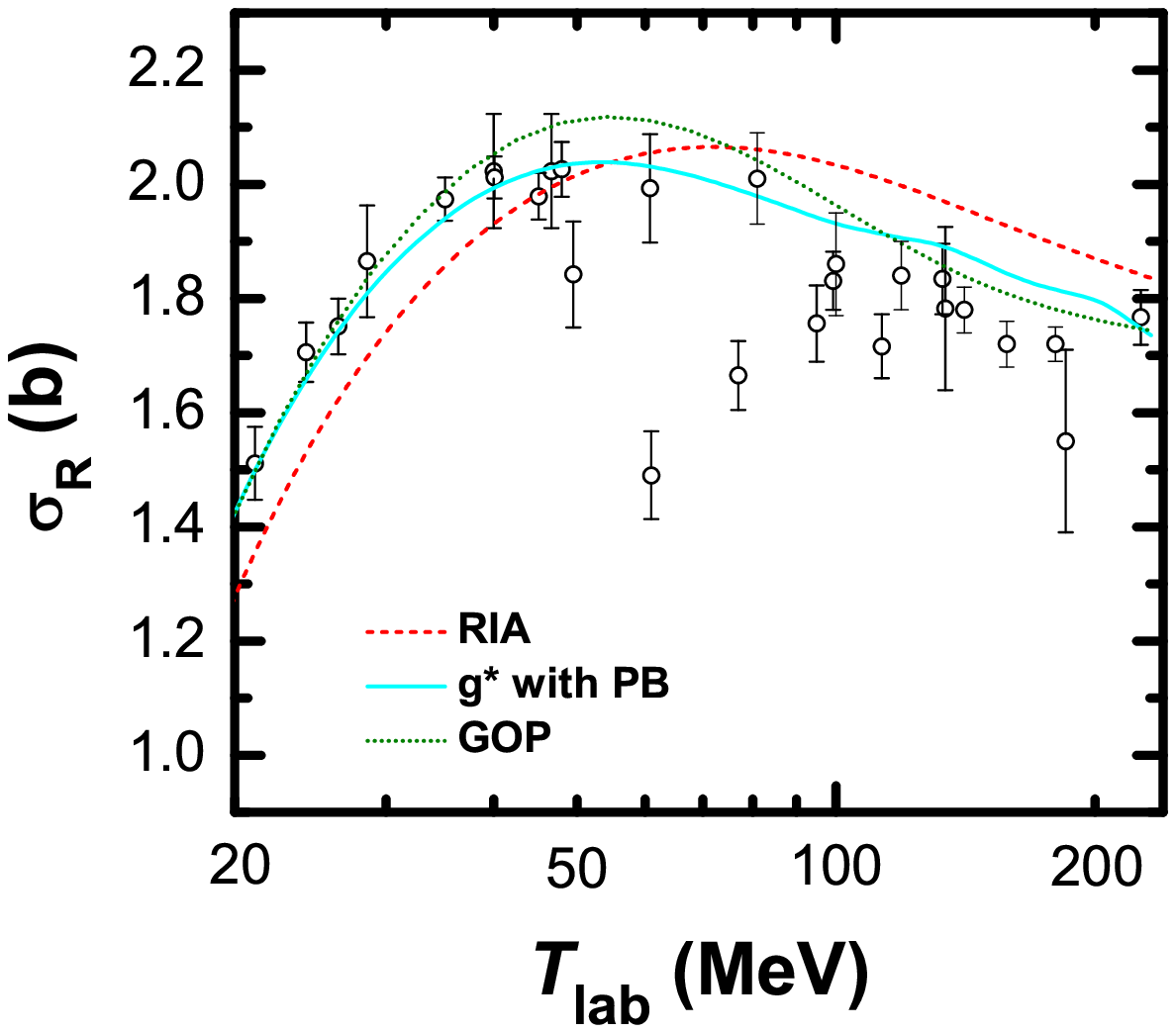}
\caption{\label{fig:sigr-Pb208}(Color online) The total reaction
cross section $\sigma_{R}$ (in units of $b$) for elastic proton
scattering from $^{208}$Pb versus the incident laboratory kinetic
energy $T_{\rm{lab}}$ (in units of MeV). The identification of the
curves is the same as in Fig.~\ref{fig:obs65-Pb208}. The
experimental data are taken from
Refs.~\cite{Go59,Me60,Go62,Ki66,Me71,Ca75,In99,Au05}.}
\end{figure}

Finally, we check the validity of the RIA model for describing total
reaction cross section $\sigma_{R}$ data for proton scattering from
$^{208}$Pb for energies lower than 200~MeV. This integral
quantity serves as crucial input in designing accelerator-driven-systems
for transmutation of radioactive waste. However, due to the
shortage of relevant $\sigma_{R}$ data, simulations are forced to
employ model-dependent values as input \cite{De01}. Furthermore,
$\sigma_{R}$ values also form an essential ingredient in studies of
nuclearsynthesis and stellar evolution. Consequently, the need for
developing predictive models of elastic proton scattering should be
evident. In Fig.~\ref{fig:sigr-Pb208} we observe that compared to the
uncorrected RIA predictions, denoted by the black dashed line
(color red online), the inclusion of both PB and density-dependent
corrections, indicated by the gray solid line (color cyan online),
provides a satisfactory description of the data: the experimental
data are from Refs.~\cite{Go59,Me60,Go62,Ki66,Me71,Ca75,In99,Au05}.
Our results are consistent with corresponding nonrelativistic
microscopic predictions \cite{De05} as well as the
GOP values (black dotted curve), over the entire energy range.  In
particular, both RIA and nonrelativistic models fail to describe the
70~MeV data: a point of concern, however, is the unsystematic trend
exhibited by the data in this region, as already mentioned in Ref.~\cite{De05}.

\section{\label{sec:summary and conclusions}Summary and conclusions}

We present the first study to examine the validity of the relativistic impulse
approximation (RIA) for describing elastic proton-nucleus scattering at
incident laboratory kinetic energies lower than 200~MeV. For simplicity
we considered a $^{208}$Pb target, which is a spin-zero spherical nucleus
for which accurately calibrated relativistic mean field nuclear structure models exist,
and for which recoil corrections to the Dirac scattering equation are expected
to be negligible. Microscopic scalar and vector optical potentials are generated
by folding our recently developed relativistic meson-exchange model of the
NN scattering matrix with appropriate Lorentz densities arising from the
PK1 Lagrangian density. We have established that phenomenological Pauli
blocking effects and density-dependent corrections
to the $\sigma$N and $\omega$N meson-nucleon coupling constants modify the RIA microscopic
scalar and vector optical potentials so as to provide a consistent and quantitative
description of all elastic scattering observables ($\sigma_{R}$, $d \sigma / d \Omega$, $A_{y}$
and $Q$) at energies ranging from 30 to 200~MeV. In particular, the effect of
PB becomes more significant at energies lower than 200~MeV, whereas phenomenological
density-dependent corrections to the NN interaction {\it also} play
an increasingly important role at energies lower than 100~MeV. Note that although
our initial study has been of a phenomenological nature, our results clearly
indicate the importance of including PB and density-dependent corrections to the
microscopic RIA model for providing consistent and quantitative predictions of
all of the above-mentioned scattering observables at energies lower than
200~MeV. Indeed the latter conclusion is consistent with corresponding nonrelativistic
microscopic studies \cite{De00b,De05}. Guided by our phenomenological result, the next
phase will be to incorporate the PB and density-dependent corrections within the
context of microscopic relativistic dynamical models in a manner similar to the
successful nonrelativistic $g$-folding model developed by the Melbourne group
\cite{Am00}.

In closing, we emphasize that not only is the RIA a highly
predictive model, but it also provides a microscopic meson-exchange
picture for understanding the behavior of the successful global
Dirac optical potentials. Another attractive feature of the RIA is
the consistent application of the relativistic meson-exchange NN
scattering matrix for generating microscopic folding optical
potentials -- required for calculating the scattering wave functions
necessary for evaluating relativistic transition matrix elements --
and also for describing the driving NN reaction mechanism in
relativistic distorted wave models of nucleon-induced reactions. One
of the most useful applications envisaged for the RIA will be to
generate microscopic optical potentials for studying elastic and
inelastic scattering of nucleons from unstable neutron- and
proton-rich nuclei for which global optical potentials do not exist.
Indeed, existing global Dirac optical potentials have been
constrained to reproduce elastic proton scattering from stable
nuclei \cite{Co93}, and hence there is no reason to believe that
these potentials can be reliably extrapolated for studies of exotic
nuclei. On the other hand, one can readily extend the RIA folding
procedure to calculate microscopic optical potentials for exotic
nuclei. The latter can easily be realized due to the current
availability of suitable relativistic meson-exchange models
\cite{Li08,Ma96,Ma98} as well sophisticated relativistic mean field
nuclear structure models for unstable nuclei \cite{To03,Me06}.
Future work will focus on studying elastic proton scattering from
exotic nuclei.

This work is partly supported by the Major State Basic Research Developing Program 2007CB815000,
the National Natural Science Foundation of China under Grant Nos. 10435010, 10775004 and 10221003,
as well as the National Research Foundation of South Africa under Grant No. 2054166.

\end{document}